\pdfoutput=1

\documentclass[%
 reprint,
 amsmath,amssymb,
 aps,prb
]{revtex4-1}

\usepackage{graphicx}
\usepackage{dcolumn}
\usepackage{bm}
\usepackage{color}

\bmdefine{\bdi}{i}
\bmdefine{\bdj}{j}
\bmdefine{\bdx}{x}
\bmdefine{\bdy}{y}
\bmdefine{\bdr}{r}
\bmdefine{\bdS}{S}
\bmdefine{\bdD}{D}
\bmdefine{\bdQ}{Q}
\bmdefine{\bdq}{q}
\bmdefine{\bdzero}{0}
\bmdefine{\bddelta}{\delta}

\begin{document}

\title{
  Effects of magnon-magnon interactions
  in a noncollinear magnet
  induced by combination of
  a symmetric and
  an antisymmetric exchange interaction
}

\author{Naoya Arakawa}
\email{naoya.arakawa@sci.toho-u.ac.jp} 
\affiliation{
Department of Physics, Toho University, 
Funabashi, Chiba, 274-8510, Japan}

\date{\today}

\begin{abstract}
  We study magnon-magnon interactions and their effects 
  in a spiral magnet induced by
  combination of an antiferromagnetic Heisenberg interaction
  and a Dzyaloshinsky-Moriya interaction. 
  We show that
  the main effect of magnon-magnon interactions on low-energy magnons
  is to renormalize the coefficient of energy dispersion.
  This could explain why some experiments
  are consistent with the noninteracting theory.
  We also show that
  although the magnon-magnon interactions induce the pair amplitude for low-energy magnons,
  its effect on the excitation energy is negligible.
  This suggests that
  for magnons the finite pair amplitude does not necessarily accompany
  the pair condensation. 

\end{abstract}
\maketitle


\section{Introduction}

Noncollinear magnets have been widely studied
in condensed matter physics.
In noncollinear magnets,  
magnetic moments form a noncollinear structure.
Some of them can be used to achieve multiferroics,
in which magnetization and electric polarization
coexist~\cite{Multiferro-review1,Multiferro-review2}.
Noncollinear magnets are also useful for realizing 
several Hall effects due to spin scalar chirality~\cite{AHE,THE}.

Some studies of frustrated noncollinear magnets 
have revealed unusual effects of magnon-magnon interactions. 
A mechanism of noncollinear magnets is
competition between symmetric exchange interactions~\cite{Yoshimori}. 
For the noncollinear magnets induced by competing Heisenberg interactions,  
the cubic terms in the magnon Hamiltonian
are finite~\cite{Miyake,Shiba,Chubukov,Cherny,RMP-Noncol},
and they cause
strong renormalization of the magnon energy 
in a wide range of the Brillouin zone~\cite{Shiba,Chubukov}
and large suppression of magnon peaks at several momenta~\cite{Cherny}.
The cubic terms are
the magnon-magnon interactions which are possible for noncollinear magnets
because they should be zero in collinear magnets~\cite{Shiba,RMP-Noncol}.
The strong renormalization and the large suppression,
which have been obtained at zero temperature, are unusual
because for collinear magnets 
magnon-magnon interactions usually have negligible effects
at sufficiently low temperatures~\cite{Dyson,Oguchi,Kanamori1,Kanamori2,Nakamura,Harris,Mano}. 

Although there is another mechanism of noncollinear magnets,
the effects of magnon-magnon interactions remain unclear.
It is combination of
a symmetric and an antisymmetric exchange interaction~\cite{Spiral-DM1};
for example,
a spiral magnetic structure can be stabilized by 
combination of
the Heisenberg and the Dzyaloshinsky-Moriya (DM) interaction~\cite{DM1,DM2,DM3} 
even without 
competing symmetric exchange interactions~\cite{Spiral-DM1,Sigrist,Spiral-DM2,Spiral-DM3}.
This mechanism can explain
noncollinear magnetic structures
in various materials (e.g., MnSi~\cite{DM-NC1,DM-NC2}, CsCuCl$_{3}$~\cite{DM-NC3},
Cr$_{1/3}$NbS$_{2}$~\cite{DM-NC4}, Ba$_{2}$CuGe$_{2}$O$_{7}$~\cite{DM-NC5}, and 
Cu$_{2}$OSeO$_{3}$~\cite{DM-NC6}).  
Although there are several studies of
interacting magnons in noncollinear magnets with a DM interaction~\cite{Frust-DM1,Frust-DM2},
their noncollinear magnetic structures are induced by competing Heisenberg interactions.
It is thus unclear how
magnon-magnon interactions affect the properties
of a noncollinear magnet induced by
combination of the Heisenberg and the DM interaction.

In this paper
we develop a theory for describing interacting magnons 
in a spiral magnet induced by combination of the Heisenberg and the DM interaction. 
Deriving its magnon Hamiltonian,
we show that 
the cubic terms become zero and
the quartic terms give the leading contributions to magnon-magnon interactions.
Then,
studying the magnon Green's function at zero temperature,
we show that
the main effect of magnon-magnon interactions
on low-energy magnons 
is to renormalize the value of the Heisenberg interaction appearing
in the coefficient of energy dispersion.
We also show that
although the magnon-magnon interactions induce the magnons-pair amplitude,
it has a negligible effect on the excitation energy. 

\section{Model}

We consider a spin Hamiltonian
\begin{align}
  H=2\sum_{\langle i,j\rangle}
  J\bdS_{i}\cdot\bdS_{j}
  +2\sum_{\langle i,j \rangle}
  \bdD_{ij}\cdot (\bdS_{i}\times\bdS_{j}),
  \label{eq:H_spin}
\end{align}
where the sum is restricted to nearest-neighbor sites. 
The first term is the antiferromagnetic Heisenberg interaction,
and the second term is the DM interaction. 

\begin{figure}
\includegraphics[width=58mm]{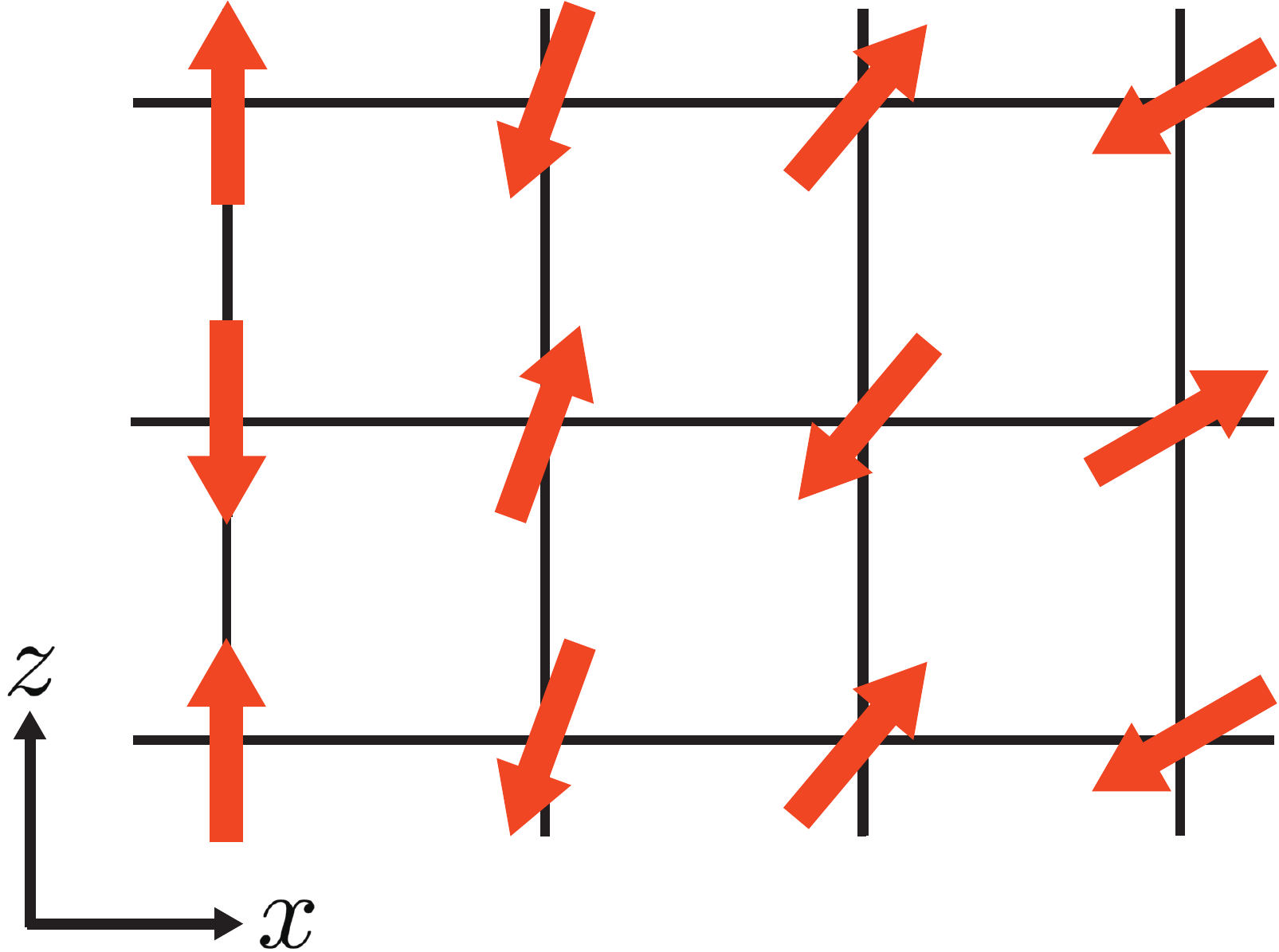}
  \caption{\label{fig1}
    Magnetic structure of our spiral magnet.
    Orange arrows represent the magnetic moments.
    The magnetic moments
    in the lines parallel to an $x$ axis
    are noncollinear due to combination of the Heisenberg and the DM interaction.
}
\end{figure}

As a concrete example,
we focus on the noncollinear magnet 
described by
\begin{align}
  \langle \bdS_{j}\rangle
  =\left(
  \begin{array}{@{\,}c@{\,}}
    S\sin \bdQ\cdot\bdr_{j}\\[3pt]
    0\\[3pt]
    S\cos \bdQ\cdot\bdr_{j}
  \end{array}
  \right).
\end{align}
Such a magnetic structure becomes the classical ground state of Eq. (\ref{eq:H_spin}),
for example, in a two-dimensional case (Fig. \ref{fig1})~\cite{Spiral-DM2}; 
in this case, 
$\bdD_{ij}$ is set to 
$\bdD_{ij}={}^{t}(0\ D_{ij}\ 0)$, where $D_{ij}=D \textrm{sgn}(x_{i}-x_{j}) \delta_{z_{i},z_{j}}$; 
and 
$\bdQ={}^{t}(Q_{x}\ Q_{z})$ is given by $Q_{x}=\pi-\tan^{-1}(D/J)$, $Q_{z}=\pi$. 
Hereafter we focus on this case. 

Before deriving the magnon Hamiltonian,
we rewrite Eq. (\ref{eq:H_spin})
in terms of different spin operators defined by
\begin{align}
  \hspace{-10pt}
  \left(
  \begin{array}{@{\,}c@{\,}}
    S^{\prime x}_{j}\\[3pt]
    S^{\prime y}_{j}\\[3pt]
    S^{\prime z}_{j}
  \end{array}
  \right)
  =
  \left(
  \begin{array}{@{\,}ccc@{\,}}
    \cos (\bdQ\cdot \bdr_{j}) & 0 & -\sin (\bdQ\cdot \bdr_{j})\\[3pt]
    0 & 1 & 0\\[3pt]
    \sin (\bdQ\cdot \bdr_{j}) & 0 & \cos (\bdQ\cdot \bdr_{j})
  \end{array}
  \right)\left(
  \begin{array}{@{\,}c@{\,}}
    S^{x}_{j}\\[3pt]
    S^{y}_{j}\\[3pt]
    S^{z}_{j}
  \end{array}
  \right).\label{eq:Rot_S}
\end{align}
Since $\bdS_{j}^{\prime}$ satisfies $\langle \bdS_{j}^{\prime}\rangle = {}^{t}(0\ 0\ S)$,
quantum fluctuations of spins can be analyzed more easily
by the spin Hamiltonian expressed in terms of $\bdS_{i}^{\prime}$ and $\bdS_{j}^{\prime}$. 
By using Eq. (\ref{eq:Rot_S}),
we can rewrite Eq. (\ref{eq:H_spin}) as follows~\cite{Spiral-DM2}
(for the details, see Appendix A):
\begin{align}
  H
  =-2\sum_{\langle i,j\rangle} 
  &\bar{J}_{ij}(S_{i}^{\prime x}S_{j}^{\prime x}+S_{i}^{\prime z}S_{j}^{\prime z})
  +2\sum_{\langle i,j\rangle}
  JS_{i}^{\prime y}S_{j}^{\prime y},
  \label{eq:H_spin-rewrite}
\end{align}
where
  \begin{align}
  \bar{J}_{ij}=
  \begin{cases}
    \sqrt{J^{2}+D^{2}} \ (|x_{i}-x_{j}|=1,\ z_{i}=z_{j}),\\
    J \ \ \ \ \ \ \ \ \ \ \  \ (|z_{i}-z_{j}|=1,\ x_{i}=x_{j}).
  \end{cases}
\end{align}

\section{Magnon Hamiltonian}

We derive
the magnon Hamiltonian including the leading terms of magnon-magnon interactions.
To express the Hamiltonian of Eq. (\ref{eq:H_spin-rewrite})
in terms of creation and annihilation operators of a magnon, 
we use
the Holstein-Primakoff
transformation~\cite{HP,NC-SW-theory1,NC-SW-theory2,NC-SW-theory3,Spiral-DM2},
\begin{align}
  S_{j}^{\prime z}=S-b^{\dagger}_{j}b_{j},
  S_{j}^{\prime -}=b^{\dagger}_{j}\sqrt{2S-b^{\dagger}_{j}b_{j}},
  S_{j}^{\prime +}=(S_{j}^{\prime -})^{\dagger},\label{eq:HP}
\end{align}
where $S_{j}^{\prime \pm}=S_{j}^{\prime x}\pm iS_{j}^{\prime y}$.
Then we expand 
$\sqrt{2S-b^{\dagger}_{j}b_{j}}$ in Eq. (\ref{eq:HP})
in powers of $b^{\dagger}_{j}b_{j}/(2S)$
and consider the leading and the next-to-leading term.
As a result, we have
\begin{align}
  &S_{j}^{\prime -}\approx \sqrt{2S}b_{j}^{\dagger}(1-\frac{1}{2}\frac{b^{\dagger}_{j}b_{j}}{2S}),\\
  &S_{j}^{\prime +}\approx \sqrt{2S}(1-\frac{1}{2}\frac{b^{\dagger}_{j}b_{j}}{2S})b_{j};
\end{align}
thus, we obtain
\begin{align}
  S_{j}^{\prime x}\approx
  \sqrt{\frac{S}{2}}[(b_{j}+b_{j}^{\dagger})
  -\frac{1}{4S}(b_{j}^{\dagger}b_{j}b_{j}+b_{j}^{\dagger}b_{j}^{\dagger}b_{j})],\label{eq:HP-S^x}\\
  S_{j}^{\prime y}\approx
  -i\sqrt{\frac{S}{2}}[(b_{j}-b_{j}^{\dagger})
  -\frac{1}{4S}(b_{j}^{\dagger}b_{j}b_{j}-b_{j}^{\dagger}b_{j}^{\dagger}b_{j})].\label{eq:HP-S^y}
\end{align}
Substituting these equations and $S_{j}^{\prime z}=S-b^{\dagger}_{j}b_{j}$
into Eq. (\ref{eq:H_spin-rewrite})
and neglecting the constant terms,
we obtain the magnon Hamiltonian $H=H_{0}+H_{\textrm{int}}$: 
$H_{0}$ represents the kinetic energy terms,
\begin{align}
  \hspace{-10pt}
  H_{0}=S\sum_{\langle i,j\rangle}\Bigl\{
  [2\bar{J}_{ij}b_{i}^{\dagger}b_{i}
    -\bar{J}_{ij}^{(+)}b_{i}b_{j}
    -\bar{J}_{ij}^{(-)}b_{i}b_{j}^{\dagger}]
  +\textrm{H.c.}\Bigr\},\label{eq:H_0}
\end{align}
where $\bar{J}_{ij}^{(\pm)}=\bar{J}_{ij}\pm J$; 
$H_{\textrm{int}}$ represents the leading terms of magnon-magnon interactions,
\begin{align}
  H_{\textrm{int}}=&\frac{1}{2}\sum_{\langle i,j\rangle}\Bigl\{
  [\bar{J}_{ij}^{(+)}b_{i}b_{j}^{\dagger}b_{j}b_{j}
    +\bar{J}_{ij}^{(-)}b_{i}b_{j}^{\dagger}b_{j}^{\dagger}b_{j}
    -2\bar{J}_{ij}b_{i}^{\dagger}b_{i}b_{j}^{\dagger}b_{j}]\notag\\
  &+\textrm{H.c.}\Bigr\}.\label{eq:H_int}
\end{align}

In our magnon Hamiltonian
the cubic terms are absent 
because the spin Hamiltonian has no terms of $S_{i}^{\prime z}S_{j}^{\prime x}$
and $S_{i}^{\prime x}S_{j}^{\prime z}$.
These terms vanish
because the contributions from the Heisenberg interaction
and from the DM interaction cancel out
(see Appendix A).
This result is distinct from 
that for
the frustrated Heisenberg models~\cite{Miyake,Shiba,Chubukov,Cherny},
in which the cubic terms are present. 

\section{Properties of noninteracting magnons}

Before analyzing
the effects of the magnon-magnon interactions,
we comment on some properties of the noninteracting magnons of our magnet.
Since
these properties are described by
the diagonalized $H_{0}$, 
we diagonalize Eq. (\ref{eq:H_0}).
It can be done by 
using the Fourier transformation and the Bogoliubov transformation~\cite{Spiral-DM2}. 
First,
by using the Fourier transforms of the quantities in Eq. (\ref{eq:H_0}),
we have
\begin{align}
  H_{0}=
  &\sum_{\bdq}
  \left(
  b^{\dagger}_{\bdq}\ b_{-\bdq}
  \right)
  \left(
  \begin{array}{@{\,}cc@{\,}}
    A_{\bdq} & B_{\bdq}\\[3pt]
    B_{\bdq} & A_{\bdq}
  \end{array}
  \right)
  \left(
  \begin{array}{@{\,}c@{\,}}
    b_{\bdq}\\[3pt]
    b_{-\bdq}^{\dagger}
  \end{array}
  \right),\label{eq:H0-mag-q}
\end{align}
where
\begin{align}
  A_{\bdq}&=S\bar{J}(\bdzero)-\frac{S}{2}\bar{J}^{(-)}(\bdq),\label{eq:A}\\  
  B_{\bdq}&=-\frac{S}{2}\bar{J}^{(+)}(\bdq),\label{eq:B}\\ 
  b_{\bdq}&=\frac{1}{\sqrt{N}}\sum_{j}b_{j}
  e^{i\bdq\cdot \bdr_{j}},\label{eq:b-Fourier}\\  
  \bar{J}(\bdq)&=\sum_{j}\bar{J}_{ij}
  e^{i\bdq\cdot (\bdr_{i}-\bdr_{j})},\label{eq:J-Fourier}\\
  \bar{J}^{(\pm)}(\bdq)&=\sum_{j}\bar{J}^{(\pm)}_{ij}
  e^{i\bdq\cdot (\bdr_{i}-\bdr_{j})}.\label{eq:J^pm-Fourier}
\end{align}
Then,
by using the Bogoliubov transformation,
\begin{align}
  \left(
  \begin{array}{@{\,}c@{\,}}
    b_{\bdq}\\[3pt]
    b_{-\bdq}^{\dagger}
  \end{array}
  \right)
  =
  \left(
  \begin{array}{@{\,}cc@{\,}}
    c_{\bdq} & -s_{\bdq}\\[3pt]
    -s_{\bdq} & c_{\bdq}
  \end{array}
  \right)
  \left(
  \begin{array}{@{\,}c@{\,}}
    \gamma_{\bdq}\\[3pt]
    \gamma_{-\bdq}^{\dagger}
  \end{array}
  \right),\label{eq:Bogo}
\end{align}
we obtain the diagonalized $H_{0}$:
\begin{align}
H_{0}=\frac{1}{2}\sum_{\bdq}\epsilon_{\bdq}
(\gamma^{\dagger}_{\bdq}\gamma_{\bdq}+\gamma_{-\bdq}\gamma_{-\bdq}^{\dagger}),
\end{align}
where
\begin{align}
  \epsilon_{\bdq}=2\sqrt{A_{\bdq}^{2}-B_{\bdq}^{2}}.
\end{align}
Note that 
$c_{\bdq}$ and $s_{\bdq}$ satisfy 
$c_{\bdq}^{2}-s_{\bdq}^{2}=1$,
$c_{\bdq}^{2}+s_{\bdq}^{2}=\frac{2A_{\bdq}}{\epsilon_{\bdq}}$,
and 
$c_{\bdq}s_{\bdq}=\frac{B_{\bdq}}{\epsilon_{\bdq}}$. 
For our spiral magnet 
$\epsilon(\bdq)$ is lowest at $\bdq=\bdzero$~\cite{Spiral-DM2},
and thus the small-$|\bdq|$ magnons describe
the low-energy excitations.

\section{Magnon Green's function}

To clarify
the effects of the magnon-magnon interactions,
we consider the magnon Green's function at zero temperature.
It can be expressed by
the $2\times 2$ matrix $G(\bdq,\omega)=(G_{ll^{\prime}})$, 
where
\begin{align}
  G_{ll^{\prime}}=\int_{-\infty}^{\infty}dt e^{i\omega t}
  (-i)\langle T x_{\bdq l}(t)x_{\bdq l^{\prime}}^{\dagger}(0)\rangle,
\end{align}
and
\begin{align}
  &x_{\bdq l}(t)=
  \begin{cases}
    e^{iH_{0}t}b_{\bdq}e^{-iH_{0}t}\ \ \ (l=1),\\
    e^{iH_{0}t}b_{-\bdq}^{\dagger}e^{-iH_{0}t}\ (l=2).
  \end{cases}\label{eq:x_ql}
\end{align}
The effects of $H_{\textrm{int}}$ can be described by
the self-energy $\Sigma(\bdq,\omega)=(\Sigma_{ll^{\prime}})$, 
\begin{align}
  \Sigma(\bdq,\omega)=G^{(0)}(\bdq,\omega)^{-1}-G(\bdq,\omega)^{-1},\label{eq:Dyson}
\end{align}
where 
\begin{align}
  \hspace{-10pt}
  G^{(0)}(\bdq,\omega)^{-1}
  =
  \left(
  \begin{array}{@{\,}cc@{\,}}
    \omega-2A_{\bdq}+i\delta & -2B_{\bdq}\\[3pt]
    -2B_{\bdq} & -\omega-2A_{\bdq}+i\delta
  \end{array}
  \right)\label{eq:G0}
\end{align}
and $\delta=0+$. 
Since 
the effects of $H_{\textrm{int}}$ can be treated perturbatively, 
it will be sufficient to calculate $\Sigma(\bdq,\omega)$ 
to first order in $H_{\textrm{int}}$ (i.e., zeroth order in $S$). 
By performing field-theoretical calculations~\cite{AGD,FW}
(for the details, see Appendix B),
we obtain
\begin{align}
  \Sigma_{11}
  =&\Sigma_{22}
  =\tilde{A}_{\bdq}
  +\frac{1}{N}\sum_{\bdq^{\prime}}\frac{B_{\bdq^{\prime}}}{\epsilon_{\bdq^{\prime}}}
  (\tilde{B}_{\bdq}+2\tilde{B}_{\bdq^{\prime}})
  \notag\\
  &+\frac{2}{N}\sum_{\bdq^{\prime}}\frac{A_{\bdq^{\prime}}}{\epsilon_{\bdq^{\prime}}}
  (\tilde{A}_{\bdq-\bdq^{\prime}}+\tilde{B}_{\bdq-\bdq^{\prime}}
  -\tilde{A}_{\bdq}-\tilde{A}_{\bdq^{\prime}}),\label{eq:Sigma_11}\\
  \Sigma_{12}
  =&\Sigma_{21}
  =\tilde{B}_{\bdq}
  -\frac{2}{N}\sum_{\bdq^{\prime}}\frac{A_{\bdq^{\prime}}}{\epsilon_{\bdq^{\prime}}}
  \tilde{B}_{\bdq}\notag\\
  &-\frac{1}{N}\sum_{\bdq^{\prime}}\frac{B_{\bdq^{\prime}}}{\epsilon_{\bdq^{\prime}}}
  (2\tilde{A}_{\bdq-\bdq^{\prime}}+2\tilde{B}_{\bdq-\bdq^{\prime}}-\tilde{A}_{\bdq}),\label{eq:Sigma_12}
\end{align}
where $\tilde{A}_{\bdq}=\frac{A_{\bdq}}{S}$ and 
$\tilde{B}_{\bdq}=\frac{B_{\bdq}}{S}$. 

We also consider $G_{\textrm{band}}(\bdq,\omega)=(G_{\nu\nu^{\prime}})$,
where 
\begin{align}
  G_{\nu\nu^{\prime}}=\int_{-\infty}^{\infty}dt e^{i\omega t}
  (-i)\langle T X_{\bdq \nu}(t)X_{\bdq \nu^{\prime}}^{\dagger}(0)\rangle
\end{align}
and
\begin{align}
  &X_{\bdq \nu}(t)=
  \begin{cases}
    e^{iH_{0}t}\gamma_{\bdq}e^{-iH_{0}t}\ \ \ (\nu=\alpha),\\
    e^{iH_{0}t}\gamma_{-\bdq}^{\dagger}e^{-iH_{0}t}\ (\nu=\beta).
  \end{cases}\label{eq:X_qnu}
\end{align}
The noninteracting $G_{\textrm{band}}(\bdq,\omega)$,
$G_{\textrm{band}}^{(0)}(\bdq,\omega)=(G^{0}_{\nu\nu^{\prime}})$, 
is given by
\begin{align}
  &G_{\alpha\alpha}^{0}=\frac{1}{\omega-\epsilon_{\bdq}+i\delta},\\
  &G_{\beta\beta}^{0}=\frac{1}{-\omega-\epsilon_{\bdq}+i\delta},\\
  &G_{\alpha\beta}^{0}=G_{\beta\alpha}^{0}=0.
\end{align}
Because of the Bogoliubov transformation, 
$G_{\nu\nu^{\prime}}$ and $G_{ll^{\prime}}$ are connected as follows: 
\begin{align}
  &G_{\alpha\alpha}
  =\tfrac{1}{2}(G_{11}-G_{22})
  +\tfrac{A_{\bdq}}{\epsilon_{\bdq}}(G_{11}+G_{22})
  +2\tfrac{B_{\bdq}}{\epsilon_{\bdq}}G_{12},\label{eq:G_alal-G_ll'}\\
  &G_{\alpha\beta}=G_{\beta\alpha}
  =2\tfrac{A_{\bdq}}{\epsilon_{\bdq}}G_{12}
  +\tfrac{B_{\bdq}}{\epsilon_{\bdq}}(G_{11}+G_{22}),\label{eq:G_albt-G_ll'}\\
  &G_{\beta\beta}
  =\tfrac{1}{2}(G_{22}-G_{11})
  +\tfrac{A_{\bdq}}{\epsilon_{\bdq}}(G_{11}+G_{22})
  +2\tfrac{B_{\bdq}}{\epsilon_{\bdq}}G_{12}.\label{eq:G_btbt-G_ll'}
\end{align}

\section{Interaction effects}

We now study
the effects of the magnon-magnon interactions
on the low-energy magnons of our magnet. 
To focus on the essential features,  
we outline the main results and implications here 
(for the details of the calculations, see Appendices C and D). 
By combining Eqs. (\ref{eq:Dyson}), (\ref{eq:G0}),
and (\ref{eq:G_alal-G_ll'}){--}(\ref{eq:G_btbt-G_ll'}),
we obtain
\begin{align}
  &G_{\alpha\alpha}\approx
  \frac{1}{\omega-\epsilon_{\bdq}^{\ast}+i\delta},\label{eq:G_alal}\\
  &G_{\beta\beta}\approx
    -\frac{1}{\omega+\epsilon_{\bdq}^{\ast}-i\delta},\label{eq:G_btbt}\\
  &G_{\alpha\beta}=G_{\beta\alpha}\approx
  \frac{g_{\bdq}}
       {(\omega-\epsilon_{\bdq}^{\ast}+i\delta)
         (\omega+\epsilon_{\bdq}^{\ast}-i\delta)},\label{eq:G_albt}
\end{align}
where
\begin{align}
  &\epsilon_{\bdq}^{\ast}=\epsilon_{\bdq}+\Delta\epsilon_{\bdq},\\
  &\Delta\epsilon_{\bdq}\approx
  \frac{2(A_{\bdq}\Sigma_{11}-B_{\bdq}\Sigma_{12})}{\epsilon_{\bdq}},\\ 
  &g_{\bdq}= \frac{2(B_{\bdq}\Sigma_{11}-A_{\bdq}\Sigma_{12})}{\epsilon_{\bdq}}\label{eq:g}.
\end{align}
For the derivations of Eqs. (\ref{eq:G_alal}){--}(\ref{eq:G_albt})
see Appendix C.

To understand the interaction effects,
we estimate $\Delta\epsilon_{\bdq}$ and $g_{\bdq}$ 
in the long-wavelength limit for $D/J \ll 1$.
In this estimate 
we introduce a cutoff momentum $q_{\textrm{c}}$, which satisfies $q_{\textrm{c}}\ll \pi$, 
and replace $\frac{1}{N}\sum_{\bdq^{\prime}}$ 
in Eqs. (\ref{eq:Sigma_11}) and (\ref{eq:Sigma_12}) 
by $\int_{0}^{q_{\textrm{c}}}\frac{dq^{\prime}}{2\pi} q^{\prime}$.
In the long-wavelength limit for $D/J \ll 1$,
$\Delta\epsilon_{\bdq}$ and $g_{\bdq}$ are given by 
\begin{align}
  &\Delta\epsilon_{\bdq}\sim 2\sqrt{2}Jq,\label{eq:DelE}\\
  &g_{\bdq}\sim -J\frac{q_{\textrm{c}}}{\pi}q,\label{eq:PairAmp}
\end{align}
as shown in Appendix D. 
Equation (\ref{eq:PairAmp}) shows that
the magnon-magnon interactions induce
the magnons-pair amplitude for finite-$q$ magnons.
Then, since $\epsilon_{\bdq}\sim 4\sqrt{2}SJq$,
we have
\begin{align}
  \epsilon_{\bdq}^{\ast}
  \sim 4\sqrt{2}SJ(1+\frac{1}{2S})q
  =4\sqrt{2}S J^{\ast}q.\label{eq:e^ast-text}
\end{align}
Since the excitation energy $\bar{\epsilon}_{\bdq}$ is determined by
\begin{align}
  \textrm{Re}[\textrm{det}G_{\textrm{band}}(\bdq,\bar{\epsilon}_{\bdq})^{-1}]=0,
\end{align}
we have
\begin{align}
  \bar{\epsilon}_{\bdq}=\sqrt{(\epsilon_{\bdq}^{\ast})^{2}-g_{\bdq}^{2}}.
\end{align}
Furthermore it reduces to
\begin{align}
  \bar{\epsilon}_{\bdq}\sim \epsilon_{\bdq}^{\ast},\label{eq:bar-e-final}
\end{align}
because
$\Delta \epsilon_{\bdq}\gg |g_{\bdq}|$ for $q_{\textrm{c}}\ll \pi$.
Note that the reason why 
  the leading correction 
  comes from the $O(S^{0})$ term is as follows: 
  in our perturbation theory, in which  
  the corrections are assumed to
  be smaller than the terms of noninteracting magnons, 
  $\bar{\epsilon}_{\bdq}$ can be written as
  $\bar{\epsilon}_{\bdq}=\sqrt{\epsilon_{\bdq}^{2}+O(S)}
  \sim \epsilon_{\bdq}[1+O(S)/(2\epsilon_{\bdq}^{2})]=\epsilon_{\bdq}+O(S^{0})$.

From these results
  we can deduce several features of the interaction effects. 
  First, Eq. (\ref{eq:bar-e-final}) shows that
  the magnons-pair amplitude, which is induced by the magnon-magnon interactions,
  has a negligible effect on the excitation energy.  
This result suggests that   
magnons-pair condensation is absent, 
although the pair amplitude is finite.
(If it occurs,
$\bar{\epsilon}_{\bdq}$ should be drastically affected by $g_{\bdq}$.)
Then Eqs. (\ref{eq:e^ast-text}) and (\ref{eq:bar-e-final}) show that 
the main effect of the magnon-magnon interactions on the low-energy magnons 
is to renormalize
the value of the Heisenberg interaction in the dispersion relation.

\section{Discussion}

We first comment on 
the absence of the magnons-pair condensation with the finite pair amplitude.  
This property is different from the property for electrons,
for which 
the finite pair amplitude accompanies the pair condensation~\cite{BCS,BCS-text}.
This difference is attributed to the following two properties. 
First, 
the magnons-pair amplitude can be induced by magnon-magnon interactions
because 
they sometimes have 
the terms that violate magnon-number conservation. 
Second, 
it is hard to realize 
the magnons-pair condensation at low temperatures 
because the description in terms of magnons
is based on the expansion in powers of $b^{\dagger}_{j}b_{j}/(2S)$. 
[Since this expansion results in 
$\epsilon_{\bdq}=O(S)$, $\Delta\epsilon_{\bdq}=O(S^{0})$, and $g_{\bdq}=O(S^{0})$,
$g_{\bdq}$ may be not large enough to induce
the pair condensation at low temperatures at which
the magnon number per site is not large.] 
Since the above discussion is applicable to more complicated magnets,
our result may be an essential property.

We then argue that 
our results provide a possible explanation of
why some experiments of noncollinear magnets
are consistent with the noninteracting theory. 
Experiments for Ba$_{2}$CuGe$_{2}$O$_{7}$~\cite{MagBand1} and Cu$_{2}$OSeO$_{3}$~\cite{MagBand2} 
showed that 
the magnon energy dispersion observed is
consistent with
that obtained in the noninteracting theory.
This is distinct from the inconsistency between experiment and
the noninteracting theory for some frustrated
noncollinear magnets~\cite{BadMagBand1-Exp,BadMagBand2-Exp,BadMagBand3-Exp1,BadMagBand3-Exp2};
such inconsistency has been proposed to
be due to the cubic terms~\cite{Shiba,BadMagBand2-Th,BandMagBand3-Th}.
Since 
Ba$_{2}$CuGe$_{2}$O$_{7}$ and Cu$_{2}$OSeO$_{3}$ 
are classified into the noncollinear magnets
induced by combination of the Heisenberg and the DM interaction~\cite{DM-NC5,DM-NC6},
the interaction effects similar to those of our magnet
could explain why
the noninteracting theory is sufficient for describing 
the magnon energy dispersion. 

We also discuss another implication of our results. 
The ratio of the DM interaction to the Heisenberg interaction for Sm$_{2}$Ir$_{2}$O$_{7}$
was estimated~\cite{MagBand3-Exp} 
by comparing some branches of magnon dispersion curves observed experimentally
with those calculated in the noninteracting theory. 
The estimated value is smaller than that derived in
the theory of superexchange interactions~\cite{MagBand3-Th}.
Since we have shown that 
in discussing the dispersion relation for low-energy magnons,  
the value of the Heisenberg interaction
changes from $J$ to $J^{\ast}(>J)$ due to the magnon-magnon interactions, 
this inconsistency could be resolved
by taking account of the similar interaction effect.

\section{Conclusions}

We have studied the magnon-magnon interactions
and their effects in the spiral magnet
induced by combination of the Heisenberg and the DM interaction.
We first derived the magnon Hamiltonian consisting of
the kinetic energy terms and the leading terms of magnon-magnon interactions.
We showed that
the cubic terms are absent
because the contributions from the Heisenberg interaction and from the DM interaction cancel out.
We then studied the magnon Green's function at zero temperature
by calculating the self-energy to first order in $H_{\textrm{int}}$.
We showed that
the main effect of the magnon-magnon interactions on the low-energy magnons 
is to replace $J$ appearing in the energy dispersion by $J^{\ast}$.
This could explain why some experiments are consistent with
the noninteracting theory
and provide a possible solution to the problem of estimating $D/J$. 
We also showed that
although the magnon-magnon interactions induce the magnons-pair amplitude,
its effect on the excitation energy is negligible.
This is attributed to the two general properties of magnons,
and it suggests that
the finite magnons-pair amplitude does not necessarily accompany
the magnons-pair condensation. 
Our results may provide a step
towards understanding the applicability of the noninteracting theory 
and interaction effects 
in various noncollinear magnets induced by combination of
a symmetric and an antisymmetric exchange interaction. 

\begin{acknowledgments}
  This work was supported by JSPS KAKENHI Grant No. JP19K14664.
\end{acknowledgments}

\appendix
\begin{widetext}
  
\section{Derivation of Eq. (\ref{eq:H_spin-rewrite})}

We derive Eq. (\ref{eq:H_spin-rewrite}).
Since this derivation has been described in Ref. \onlinecite{Spiral-DM2},
we outline it here.
Substituting Eq. (\ref{eq:Rot_S}) into Eq. (\ref{eq:H_spin}),
we obtain
\begin{align}
  H
  =&2\sum_{\langle i,j\rangle}J\cos[\bdQ\cdot (\bdr_{i}-\bdr_{j})]
  (S_{i}^{\prime x}S_{j}^{\prime x}+S_{i}^{\prime z}S_{j}^{\prime z})
  +2\sum_{\langle i,j\rangle}J\sin[\bdQ\cdot (\bdr_{i}-\bdr_{j})]
  (S_{i}^{\prime z}S_{j}^{\prime x}-S_{i}^{\prime x}S_{j}^{\prime z})
  +2\sum_{\langle i,j\rangle}JS_{i}^{\prime y}S_{j}^{\prime y}\notag\\
  &-2\sum_{\langle i,j\rangle}D_{ij}\sin[\bdQ\cdot (\bdr_{i}-\bdr_{j})]
  (S_{i}^{\prime x}S_{j}^{\prime x}+S_{i}^{\prime z}S_{j}^{\prime z})
  +2\sum_{\langle i,j\rangle}D_{ij}\cos[\bdQ\cdot (\bdr_{i}-\bdr_{j})]
  (S_{i}^{\prime z}S_{j}^{\prime x}-S_{i}^{\prime x}S_{j}^{\prime z})\notag\\ 
  =&2\sum_{\langle i,j\rangle}M_{ij}
  (S_{i}^{\prime x}S_{j}^{\prime x}+S_{i}^{\prime z}S_{j}^{\prime z})
  +2\sum_{\langle i,j\rangle}N_{ij}
  (S_{i}^{\prime z}S_{j}^{\prime x}-S_{i}^{\prime x}S_{j}^{\prime z})
  +2\sum_{\langle i,j\rangle}JS_{i}^{\prime y}S_{j}^{\prime y},\label{eq:Hspin-rewrite1}
\end{align}
where
\begin{align}
  M_{ij}=J\cos[\bdQ\cdot (\bdr_{i}-\bdr_{j})]
  -D_{ij}\sin[\bdQ\cdot (\bdr_{i}-\bdr_{j})],\label{eq:M}\\
  N_{ij}=J\sin[\bdQ\cdot (\bdr_{i}-\bdr_{j})]
  +D_{ij}\cos[\bdQ\cdot (\bdr_{i}-\bdr_{j})].\label{eq:N}
\end{align}
Note that
for our magnet 
$D_{ij}$ is given by 
\begin{align}
  &D_{ij}=D \textrm{sgn}(x_{i}-x_{j}) \delta_{z_{i},z_{j}},\label{eq:DM}
\end{align}
and $\bdQ={}^{t}(Q_{x}\ Q_{z})$ is given by 
\begin{align}
  &\cos Q_{x}=-\frac{J}{\sqrt{J^{2}+D^{2}}},\ \sin Q_{x}=\frac{D}{\sqrt{J^{2}+D^{2}}},\label{eq:Q_x}\\
  &\cos Q_{z}=-1,\ \sin Q_{z}=0.\label{eq:Q_z}
\end{align}
By using Eqs. (\ref{eq:DM}){--}(\ref{eq:Q_z}), 
we can write $M_{ij}$ and $N_{ij}$ as follows:
for $\bdr_{i}-\bdr_{j}={}^{t}(x_{i}-x_{j}\ z_{i}-z_{j})={}^{t}(\pm 1\ 0)$
they are 
\begin{align}
  &M_{ij}=J\cos Q_{x}-D\sin Q_{x}=-\sqrt{J^{2}+D^{2}},\label{eq:M-x}\\
  &N_{ij}=J\sin Q_{x}+D\cos Q_{x}=0,\label{eq:N-x}
\end{align}
while for $\bdr_{i}-\bdr_{j}={}^{t}(x_{i}-x_{j}\ z_{i}-z_{j})={}^{t}(0\ \pm 1)$
they are
\begin{align}
  &M_{ij}=-J,\label{eq:M-z}\\
  &N_{ij}=0.\label{eq:N-z}
\end{align}
Combining Eqs. (\ref{eq:M-x}){--}(\ref{eq:N-z}) with Eq. (\ref{eq:Hspin-rewrite1}),
we obtain Eq. (\ref{eq:H_spin-rewrite}). 

The above argument shows that
the terms of $S_{i}^{\prime z}S_{j}^{\prime x}$ and $S_{i}^{\prime x}S_{j}^{\prime z}$ 
disappear in Eq. (\ref{eq:H_spin-rewrite})
because $N_{ij}$ becomes zero, i.e.,
the contributions from the Heisenberg interaction and from the DM interaction
cancel out [see Eq. (\ref{eq:N-x})]. 

\section{Derivation of Eqs. (\ref{eq:Sigma_11}) and (\ref{eq:Sigma_12})}

We derive Eqs. (\ref{eq:Sigma_11}) and (\ref{eq:Sigma_12}).
This derivation consists of three steps. 
(It is similar to
the derivation of the self-energy of an electron
in the Hartree-Fock approximation~\cite{FW}.)

By treating $H_{\textrm{int}}$ perturbatively, 
we can express our magnon Green's function
$G(\bdq,\omega)=(G_{ll^{\prime}})$ as follows:
\begin{align}
  G_{ll^{\prime}}
  =&\int_{-\infty}^{\infty}dt e^{i\omega t}
  (-i)\langle T x_{\bdq l}(t)x_{\bdq l^{\prime}}^{\dagger}(0)\rangle\notag\\
  =&\int_{-\infty}^{\infty}dt e^{i\omega t}
  (-i)\langle T x_{\bdq l}(t)x_{\bdq l^{\prime}}^{\dagger}(0)\rangle_{0}
  +\int_{-\infty}^{\infty}dt e^{i\omega t}\int_{-\infty}^{\infty}dt_{1} 
  (-i)^{2}\langle T H_{\textrm{int}}(t_{1}) x_{\bdq l}(t)x_{\bdq l^{\prime}}^{\dagger}(0)\rangle_{0}
  +\cdots,\label{eq:G-perturb}
\end{align}
where $x_{\bdq l}(t)$ is given by Eq. (\ref{eq:x_ql}), 
and $H_{\textrm{int}}(t_{1})$ is given by
\begin{align}
  &H_{\textrm{int}}(t_{1})=e^{iH_{0}t_{1}}H_{\textrm{int}}e^{-iH_{0}t_{1}}.\label{eq:H_int^I}
\end{align}
Note that in Eq. (\ref{eq:G-perturb})
the subscript $0$ indicates the evaluation in the unperturbed state.

Before calculating the self-energy, we rewrite $H_{\textrm{int}}$. 
By using the Fourier transforms of the quantities
appearing in Eq. (\ref{eq:H_int}),
we can write $H_{\textrm{int}}$ as follows:
\begin{align}
  H_{\textrm{int}}
  =\sum_{\bdq_{1},\bdq_{2},\bdq_{3},\bdq_{4}}
  V^{(\textrm{a})}(\bdq_{1},\bdq_{2},\bdq_{3},\bdq_{4})
  (b_{\bdq_{1}}^{\dagger}b_{\bdq_{2}}b_{\bdq_{3}}b_{\bdq_{4}}
  +b_{\bdq_{4}}^{\dagger}b_{\bdq_{3}}^{\dagger}b_{\bdq_{2}}^{\dagger}b_{\bdq_{1}})
  +\sum_{\bdq_{1},\bdq_{2},\bdq_{3},\bdq_{4}}
  V^{(\textrm{b})}(\bdq_{1},\bdq_{2},\bdq_{3},\bdq_{4})
  b_{\bdq_{1}}^{\dagger}b_{\bdq_{2}}^{\dagger}b_{\bdq_{3}}b_{\bdq_{4}},\label{eq:H_int-mom}
\end{align}
where
\begin{align}
  V^{(\textrm{a})}(\bdq_{1},\bdq_{2},\bdq_{3},\bdq_{4})
  &=\frac{1}{12N}\delta_{\bdq_{1},\bdq_{2}+\bdq_{3}+\bdq_{4}}
  [\bar{J}^{(+)}(\bdq_{2})+\bar{J}^{(+)}(\bdq_{3})+\bar{J}^{(+)}(\bdq_{4})]\label{eq:V^a-pre},\\
  V^{(\textrm{b})}(\bdq_{1},\bdq_{2},\bdq_{3},\bdq_{4})
  &=\frac{1}{8N}\delta_{\bdq_{1}+\bdq_{2},\bdq_{3}+\bdq_{4}}
  [\bar{J}^{(-)}(\bdq_{1})+\bar{J}^{(-)}(\bdq_{2})
    +\bar{J}^{(-)}(\bdq_{3})+\bar{J}^{(-)}(\bdq_{4})]\notag\\
  &-\frac{1}{4N}\delta_{\bdq_{1}+\bdq_{2},\bdq_{3}+\bdq_{4}}
  [\bar{J}(\bdq_{1}-\bdq_{3})+\bar{J}(\bdq_{1}-\bdq_{4})
    +\bar{J}(\bdq_{2}-\bdq_{3})+\bar{J}(\bdq_{2}-\bdq_{4})]{\color{red}.}\label{eq:V^b-pre}
\end{align}
Note that $b_{\bdq}$, $\bar{J}(\bdq)$, and $\bar{J}^{(\pm)}(\bdq)$
are given by Eqs. (\ref{eq:b-Fourier}), (\ref{eq:J-Fourier}),
and (\ref{eq:J^pm-Fourier}), respectively.

We now calculate the self-energy to first order in $H_{\textrm{int}}$. 
By using Eqs. (\ref{eq:G-perturb}) and (\ref{eq:H_int-mom}) and the Wick's theorem
and performing the field-theoretical calculations~\cite{FW,AGD},
we obtain
\begin{align}
  \Sigma_{11}
  =\Sigma_{11}^{(\textrm{a}1)}+\Sigma_{11}^{(\textrm{a}2)}+\Sigma_{11}^{(\textrm{b})},\label{eq:Sig11-start}
\end{align}
where $\Sigma_{11}^{(\textrm{a}1)}$ and $\Sigma_{11}^{(\textrm{a}2)}$
are the self-energy terms
due to the $V^{(\textrm{a})}$ terms in Eq. (\ref{eq:H_int-mom}),
\begin{align}
  \Sigma_{11}^{(\textrm{a}1)}
  =&\sum_{\bdq^{\prime}}
  [V^{(\textrm{a})}(\bdq,\bdq,\bdq^{\prime},-\bdq^{\prime})
  +V^{(\textrm{a})}(\bdq,\bdq^{\prime},\bdq,-\bdq^{\prime})
  +V^{(\textrm{a})}(\bdq,\bdq^{\prime},-\bdq^{\prime},\bdq)]
  \langle b_{\bdq^{\prime}}b_{-\bdq^{\prime}}\rangle_{0},\\
  \Sigma_{11}^{(\textrm{a}2)}
  =&\sum_{\bdq^{\prime}}
  [V^{(\textrm{a})}(\bdq,\bdq,\bdq^{\prime},-\bdq^{\prime})
  +V^{(\textrm{a})}(\bdq,\bdq^{\prime},\bdq,-\bdq^{\prime})
  +V^{(\textrm{a})}(\bdq,\bdq^{\prime},-\bdq^{\prime},\bdq)]
  \langle b_{-\bdq^{\prime}}^{\dagger}b_{\bdq^{\prime}}^{\dagger}\rangle_{0},
\end{align}
and $\Sigma_{11}^{(\textrm{b})}$ is the self-energy term
due to the $V^{(\textrm{b})}$ term in Eq. (\ref{eq:H_int-mom}),
\begin{align}
  \Sigma_{11}^{(\textrm{b})}
  =\sum_{\bdq^{\prime}}
  [V^{(\textrm{b})}(\bdq,\bdq^{\prime},\bdq,\bdq^{\prime})
  +V^{(\textrm{b})}(\bdq,\bdq^{\prime},\bdq^{\prime},\bdq)
  +V^{(\textrm{b})}(\bdq^{\prime},\bdq,\bdq,\bdq^{\prime})
  +V^{(\textrm{b})}(\bdq^{\prime},\bdq,\bdq^{\prime},\bdq)]
  \langle b_{\bdq^{\prime}}^{\dagger}b_{\bdq^{\prime}}\rangle_{0}.
\end{align}
Because of the Bogoliubov transformation
Eq. (\ref{eq:Bogo}),
we have
\begin{align}
  &\langle b_{\bdq^{\prime}}b_{-\bdq^{\prime}}\rangle_{0}
  =\langle b_{-\bdq^{\prime}}^{\dagger}b_{\bdq^{\prime}}^{\dagger}\rangle_{0}
  =-s_{\bdq^{\prime}}c_{\bdq^{\prime}},\label{eq:bb}\\
  &\langle b_{\bdq^{\prime}}^{\dagger}b_{\bdq^{\prime}}\rangle_{0}
  =s_{\bdq^{\prime}}^{2}.\label{eq:b*b}
\end{align}
Since $c_{\bdq}$ and $s_{\bdq}$ satisfy 
$c_{\bdq}^{2}-s_{\bdq}^{2}=1$ and
$c_{\bdq}^{2}+s_{\bdq}^{2}=\frac{2A_{\bdq}}{\epsilon_{\bdq}}$,
where $\epsilon_{\bdq}=2\sqrt{A_{\bdq}^{2}-B_{\bdq}^{2}}$, 
Eqs. (\ref{eq:bb}) and (\ref{eq:b*b}) become
\begin{align}
  &\langle b_{\bdq^{\prime}}b_{\bdq^{\prime}}\rangle_{0}
  =\langle b_{\bdq^{\prime}}^{\dagger}b_{\bdq^{\prime}}^{\dagger}\rangle_{0}
  =-\frac{B_{\bdq^{\prime}}}{\epsilon_{\bdq^{\prime}}},\label{eq:bb_rew}\\
  &\langle b_{\bdq^{\prime}}^{\dagger}b_{\bdq^{\prime}}\rangle_{0}
  =-\frac{1}{2}+\frac{A_{\bdq^{\prime}}}{\epsilon_{\bdq^{\prime}}}.\label{eq:b*b_rew}
\end{align}
Combining Eqs. (\ref{eq:Sig11-start}){--}(\ref{eq:b*b_rew})
with Eqs. (\ref{eq:V^a-pre}) and (\ref{eq:V^b-pre}),
we obtain
\begin{align}
  \Sigma_{11}
  =&\frac{1}{2N}\sum_{\bdq^{\prime}}
  [2\bar{J}(\bdzero)+2\bar{J}(\bdq-\bdq^{\prime})
   -\bar{J}^{(-)}(\bdq)-\bar{J}^{(-)}(\bdq^{\prime})]
  -\frac{1}{N}\sum_{\bdq^{\prime}}
  [2\bar{J}(\bdzero)+2\bar{J}(\bdq-\bdq^{\prime})
   -\bar{J}^{(-)}(\bdq)-\bar{J}^{(-)}(\bdq^{\prime})]
    \frac{A_{\bdq^{\prime}}}{\epsilon_{\bdq^{\prime}}}\notag\\
  &-\frac{1}{2N}\sum_{\bdq^{\prime}}
   [\bar{J}^{(+)}(\bdq)+2\bar{J}^{(+)}(\bdq^{\prime})]
   \frac{B_{\bdq^{\prime}}}{\epsilon_{\bdq^{\prime}}}.\label{eq:Sig11}      
\end{align}
In a similar way
we can calculate $\Sigma_{12}$, $\Sigma_{21}$, and $\Sigma_{22}$; 
the results are 
\begin{align}
  \Sigma_{12}
  =&-\frac{1}{4N}\sum_{\bdq^{\prime}}
    [2\bar{J}^{(+)}(\bdq)+\bar{J}^{(+)}(\bdq^{\prime})]
    +\frac{1}{2N}\sum_{\bdq^{\prime}}
    [2\bar{J}^{(+)}(\bdq)+\bar{J}^{(+)}(\bdq^{\prime})]
        \frac{A_{\bdq^{\prime}}}{\epsilon_{\bdq^{\prime}}}\notag\\
    &+\frac{1}{2N}\sum_{\bdq^{\prime}}
    [2\bar{J}(\bdq-\bdq^{\prime})+2\bar{J}(\bdq+\bdq^{\prime})
     -\bar{J}^{(-)}(\bdq)-\bar{J}^{(-)}(\bdq^{\prime})]
    \frac{B_{\bdq^{\prime}}}{\epsilon_{\bdq^{\prime}}},\label{eq:Sig12}\\
  \Sigma_{21}=&\Sigma_{12},\\
  \Sigma_{22}=&\Sigma_{11}.\label{eq:Sig22}
\end{align}
Since $\bar{J}^{(+)}(\bdq)$, $\bar{J}^{(-)}(\bdq)$, and $\bar{J}(\bdq)$ satisfy
\begin{align}
  \bar{J}^{(+)}(\bdq)=-2\tilde{B}_{\bdq},\
  \bar{J}^{(-)}(\bdq)=2\bar{J}(\bdzero)-2\tilde{A}_{\bdq},\
  \bar{J}(\bdq)=\bar{J}(\bdzero)-\tilde{A}_{\bdq}-\tilde{B}_{\bdq},
\end{align}
where $\tilde{A}_{\bdq}=\frac{A_{\bdq}}{S}$ and $\tilde{B}_{\bdq}=\frac{B_{\bdq}}{S}$,
we can rewrite 
Eqs. (\ref{eq:Sig11}){--}(\ref{eq:Sig22}) as follows:
\begin{align}
  \Sigma_{11}
  =\Sigma_{22}
  =&-\frac{1}{N}\sum_{\bdq^{\prime}}
    (\tilde{A}_{\bdq-\bdq^{\prime}}+\tilde{B}_{\bdq-\bdq^{\prime}}
    -\tilde{A}_{\bdq}-\tilde{A}_{\bdq^{\prime}})
   +\frac{2}{N}\sum_{\bdq^{\prime}}
    (\tilde{A}_{\bdq-\bdq^{\prime}}+\tilde{B}_{\bdq-\bdq^{\prime}}
    -\tilde{A}_{\bdq}-\tilde{A}_{\bdq^{\prime}})
    \frac{A_{\bdq^{\prime}}}{\epsilon_{\bdq^{\prime}}}\notag\\
   &+\frac{1}{N}\sum_{\bdq^{\prime}}
    (\tilde{B}_{\bdq}+2\tilde{B}_{\bdq^{\prime}})
    \frac{B_{\bdq^{\prime}}}{\epsilon_{\bdq^{\prime}}},\label{eq:Sig11-rewrite}\\
  \Sigma_{12}
  =\Sigma_{21}
  =&\frac{1}{2N}\sum_{\bdq^{\prime}}
   (2\tilde{B}_{\bdq}+\tilde{B}_{\bdq^{\prime}})
   -\frac{1}{N}\sum_{\bdq^{\prime}}
    (2\tilde{B}_{\bdq}+\tilde{B}_{\bdq^{\prime}})
    \frac{A_{\bdq^{\prime}}}{\epsilon_{\bdq^{\prime}}}
   -\frac{1}{N}\sum_{\bdq^{\prime}}
    (2\tilde{A}_{\bdq-\bdq^{\prime}}+2\tilde{B}_{\bdq-\bdq^{\prime}}
    -\tilde{A}_{\bdq}-\tilde{A}_{\bdq^{\prime}})
    \frac{B_{\bdq^{\prime}}}{\epsilon_{\bdq^{\prime}}}.\label{eq:Sig12-rewrite}
\end{align}
Then, since $\tilde{A}_{\bdq}$ and $\tilde{B}_{\bdq}$ satisfy
\begin{align}
  &\sum_{\bdq^{\prime}}\tilde{A}_{\bdq-\bdq^{\prime}}-\sum_{\bdq^{\prime}}\tilde{A}_{\bdq^{\prime}}=0,\\
  &\sum_{\bdq^{\prime}}\tilde{B}_{\bdq-\bdq^{\prime}}=0,
\end{align}
Eqs. (\ref{eq:Sig11-rewrite}) and (\ref{eq:Sig12-rewrite}) 
reduce to Eqs. (\ref{eq:Sigma_11}) and (\ref{eq:Sigma_12}). 

\section{Derivation of Eqs. (\ref{eq:G_alal}){--}(\ref{eq:G_albt})}

We derive Eqs. (\ref{eq:G_alal}){--}(\ref{eq:G_albt}). 
This derivation consists of two steps.

First, we derive the expression of $G(\bdq,\omega)=(G_{ll^{\prime}})$. 
Combining Eqs. (\ref{eq:Dyson}) and (\ref{eq:G0}),
we obtain
\begin{align}
  G(\bdq,\omega)
  =\frac{1}{(-\textrm{det}G^{-1})}
  \left(
  \begin{array}{@{\,}cc@{\,}}
    \omega-i\delta+2A_{\bdq}+\Sigma_{11} & -2B_{\bdq}-\Sigma_{12}\\[3pt]
    -2B_{\bdq}-\Sigma_{12} & -\omega-i\delta+2A_{\bdq}+\Sigma_{11}
  \end{array}
  \right),\label{eq:G}
\end{align}
where
\begin{align}
  -\textrm{det}G^{-1}
  =&(\omega-i\delta+2A_{\bdq}+\Sigma_{11})(\omega+i\delta-2A_{\bdq}-\Sigma_{11})
  +(2B_{\bdq}+\Sigma_{12})^{2}\notag\\
  =&(\omega-i\delta)(\omega+i\delta)-(\epsilon_{\bdq}^{\ast})^{2}\notag\\
  =&(\omega+\epsilon_{\bdq}^{\ast}-i\delta)(\omega-\epsilon_{\bdq}^{\ast}+i\delta),
  \label{eq:G^-1}
\end{align}
and
\begin{align}
  \epsilon_{\bdq}^{\ast}
  =&\sqrt{(2A_{\bdq}+\Sigma_{11})^{2}-(2B_{\bdq}+\Sigma_{12})^{2}}\notag\\
  =&\sqrt{[2(A_{\bdq}+B_{\bdq})+(\Sigma_{11}+\Sigma_{12})]
    [2(A_{\bdq}-B_{\bdq})+(\Sigma_{11}-\Sigma_{12})]}
  \notag\\
  =&\epsilon_{\bdq}
  \Bigl[1
    +\frac{4(A_{\bdq}\Sigma_{11}-B_{\bdq}\Sigma_{12})+(\Sigma_{11}^{2}-\Sigma_{12}^{2})}
    {\epsilon_{\bdq}^{2}}\Bigr]^{\frac{1}{2}}.\label{eq:e^ast-full}
\end{align}
In deriving the third line of Eq. (\ref{eq:e^ast-full})
we have used $\epsilon_{\bdq}^{2}=4(A_{\bdq}^{2}-B_{\bdq}^{2})$.
Since we have calculated the self-energy perturbatively, 
we may write Eq. (\ref{eq:e^ast-full}) as follows:
\begin{align}
  \epsilon_{\bdq}^{\ast}
  \approx &
  \epsilon_{\bdq}
  +\frac{2(A_{\bdq}\Sigma_{11}-B_{\bdq}\Sigma_{12})}{\epsilon_{\bdq}}
  =\epsilon_{\bdq}+\Delta\epsilon_{\bdq}.\label{eq:e^ast-approx}
\end{align}
In deriving Eq. (\ref{eq:e^ast-approx})
  from Eq. (\ref{eq:e^ast-full})
  we have omitted the term $(\Sigma_{11}^{2}-\Sigma_{12}^{2})$
  because 
  it is of higher order in $S^{-1}$ than the term $4(A_{\bdq}\Sigma_{11}-B_{\bdq}\Sigma_{12})$;
  the former and the latter are $O(S^{0})$ and $O(S^{1})$, respectively.
  This procedure is necessary for consistency within the perturbation theory
  in which
  the expansion in powers of
  a ratio of the magnon number operator, such as $b_{j}^{\dagger}b_{j}$, to $(2S)$ is used 
  because the omitted term is of the same order as
  a product of $A_{\bdq}$ (or $B_{\bdq}$) and the self-energy
  due to the second-order terms of $H_{\textrm{int}}$. 

Then we combine Eq. (\ref{eq:G}) with
Eqs. (\ref{eq:G_alal-G_ll'}){--}(\ref{eq:G_btbt-G_ll'}).
Since Eq. (\ref{eq:G}) shows
\begin{align}
  G_{11}-G_{22}&=\frac{2}{(-\textrm{det}G^{-1})}\omega,\\
  G_{11}+G_{22}&=\frac{2}{(-\textrm{det}G^{-1})}(2A_{\bdq}+\Sigma_{11}-i\delta),\\
  2G_{12}&=-\frac{2}{(-\textrm{det}G^{-1})}(2B_{\bdq}+\Sigma_{12}),
\end{align}
we can express Eq. (\ref{eq:G_alal-G_ll'}) as follows:
\begin{align}
  G_{\alpha\alpha}
  =&\frac{1}{(-\textrm{det}G^{-1})}
  \Bigl[\omega+\frac{2A_{\bdq}}{\epsilon_{\bdq}}(2A_{\bdq}+\Sigma_{11}-i\delta)
    -\frac{2B_{\bdq}}{\epsilon_{\bdq}}(2B_{\bdq}+\Sigma_{12})\Bigr]\notag\\
  \approx &
  \frac{\omega+\epsilon^{\ast}_{\bdq}-i\delta}
       {(\omega+\epsilon_{\bdq}^{\ast}-i\delta)(\omega-\epsilon_{\bdq}^{\ast}+i\delta)}\notag\\
   =&\frac{1}{\omega-\epsilon_{\bdq}^{\ast}+i\delta}.\label{eq:G_alpalp}
\end{align}
In a similar way
we obtain
\begin{align}
  G_{\beta\beta}
  \approx &
  \frac{-\omega+\epsilon^{\ast}_{\bdq}-i\delta}
       {(\omega+\epsilon_{\bdq}^{\ast}-i\delta)(\omega-\epsilon_{\bdq}^{\ast}+i\delta)}\notag\\
  =&-\frac{1}{\omega+\epsilon_{\bdq}^{\ast}-i\delta},\label{eq:G_betbet}
\end{align}
and
\begin{align}
  G_{\alpha\beta}
  =&\frac{1}{(-\textrm{det}G^{-1})}
  \Bigl(\frac{2B_{\bdq}}{\epsilon_{\bdq}}\Sigma_{11}
    -\frac{2A_{\bdq}}{\epsilon_{\bdq}}\Sigma_{12}\Bigr)\notag\\
  =&\frac{g_{\bdq}}
  {(\omega+\epsilon_{\bdq}^{\ast}-i\delta)(\omega-\epsilon_{\bdq}^{\ast}+i\delta)},
  \label{eq:G_alpbet}
\end{align}
where $g_{\bdq}$ is given by Eq. (\ref{eq:g}). 

\section{Derivation of Eqs. (\ref{eq:DelE}) and (\ref{eq:PairAmp})}

We derive Eqs. (\ref{eq:DelE}) and (\ref{eq:PairAmp})
by estimating $\Delta\epsilon_{\bdq}$ and $g_{\bdq}$
in the long-wavelength limit for $D/J \ll 1$.

We first calculate $\Delta\epsilon_{\bdq}$
in the long-wavelength limit for $D/J \ll 1$. 
From Eq. (\ref{eq:e^ast-approx}) and
Eqs. (\ref{eq:Sigma_11}) and (\ref{eq:Sigma_12}), 
we have
\begin{align}
  \Delta\epsilon_{\bdq}
  =\frac{1}{2S}\epsilon_{\bdq}
  +\frac{2}{N}\sum_{\bdq^{\prime}}
  (\tilde{A}_{\bdq-\bdq^{\prime}}+\tilde{A}_{\bdq+\bdq^{\prime}}
  +\tilde{B}_{\bdq-\bdq^{\prime}}+\tilde{B}_{\bdq+\bdq^{\prime}})
  \Bigl(\frac{A_{\bdq}A_{\bdq^{\prime}}+B_{\bdq}B_{\bdq^{\prime}}}
       {\epsilon_{\bdq}\epsilon_{\bdq^{\prime}}}\Bigr)
  -\frac{1}{N}\sum_{\bdq^{\prime}}
  \Bigl(\epsilon_{\bdq}\frac{\tilde{A}_{\bdq^{\prime}}}{\epsilon_{\bdq^{\prime}}}
  +\frac{\tilde{A}_{\bdq}}{\epsilon_{\bdq}}
  \epsilon_{\bdq^{\prime}}\Bigr).\label{eq:DelE-start}
\end{align}
In the long-wavelength limit for $D/J\ll 1$, 
Eq. (\ref{eq:DelE-start}) becomes
\begin{align}
  \Delta\epsilon_{\bdq}
  \sim
  2\sqrt{2}Jq
  +\frac{1}{N}\sum_{\bdq^{\prime}}(q^{2}+q^{\prime 2})\frac{\tilde{A}_{\bdzero}^{2}}{4Jqq^{\prime}}
  -\frac{1}{N}\tilde{A}_{\bdzero}\sum_{\bdq^{\prime}}
  (\frac{q}{q^{\prime}}+\frac{q^{\prime}}{q}),\label{eq:DelE-next}
\end{align}
where we have used
\begin{align}
  \epsilon_{\bdq}
  &=4S[J^{2}(2-\cos q_{x}\cos q_{z}-\cos^{2} q_{z})
    +J\sqrt{J^{2}+D^{2}}(2-\cos^{2}q_{x}-\cos q_{x}\cos q_{z})
    +D^{2}(1-\cos q_{x})]^{\frac{1}{2}}\notag\\
  &\sim 4\sqrt{2}S J q,\label{eq:e-limit}\\
  \tilde{A}_{\bdq}
  &=2\sqrt{J^{2}+D^{2}}+2J-(\sqrt{J^{2}+D^{2}}-J)\cos q_{x}\notag\\
  &\sim 4J+\frac{D^{2}}{2J}\notag\\
  &=\tilde{A}_{\bdzero},\label{eq:A-limit}\\
  \tilde{B}_{\bdq}
  &=-(\sqrt{J^{2}+D^{2}}+J)\cos q_{x}-2J\cos q_{z}\notag\\
  &\sim -4J-\frac{D^{2}}{2J}+Jq^{2}\notag\\
  &=-\tilde{A}_{\bdzero}+Jq^{2}.\label{eq:B-limit}
\end{align}
In addition,
since $\tilde{A}_{\bdzero}^{2}=(4J+\frac{D^{2}}{2J})^{2}\sim 16J^{2}+4D^{2}$,
we obtain 
\begin{align}
  \Delta\epsilon_{\bdq}
  &\sim 2\sqrt{2}Jq
  +\frac{1}{2N}\frac{D^{2}}{J}\sum_{\bdq^{\prime}}(\frac{q}{q^{\prime}}+\frac{q^{\prime}}{q})\notag\\
  &\sim 2\sqrt{2}Jq+\frac{1}{3}\frac{D^{2}}{J^{2}}\frac{q_{\textrm{c}}}{\pi}Jq_{\textrm{c}}\notag\\
  &\sim 2\sqrt{2}Jq.\label{eq:DelE-last}
\end{align}
In deriving the second line of Eq. (\ref{eq:DelE-last})
we have
calculated the second term on the right-hand side
by replacing $\frac{1}{N}\sum_{\bdq^{\prime}}$ by
$\int_{0}^{q_{\textrm{c}}}\frac{dq^{\prime}}{2\pi}q^{\prime}$
and approximating the value of $q$ by $q_{\textrm{c}}$.

In a similar way we can calculate $g_{\bdq}$
in the long-wavelength limit for $D/J \ll 1$.
From Eq. (\ref{eq:g}) and
Eqs. (\ref{eq:Sigma_11}) and (\ref{eq:Sigma_12}), 
we have
\begin{align}
  g_{\bdq}=\frac{2}{N}\sum_{\bdq^{\prime}}
  (\tilde{A}_{\bdq-\bdq^{\prime}}+\tilde{A}_{\bdq+\bdq^{\prime}}
  +\tilde{B}_{\bdq-\bdq^{\prime}}+\tilde{B}_{\bdq+\bdq^{\prime}})
  \Bigl(\frac{A_{\bdq}B_{\bdq^{\prime}}+B_{\bdq}A_{\bdq^{\prime}}}
       {\epsilon_{\bdq}\epsilon_{\bdq^{\prime}}}\Bigr)
       -\frac{1}{N}\sum_{\bdq^{\prime}}
       \Bigl(\frac{\tilde{B}_{\bdq}}{\epsilon_{\bdq}}\epsilon_{\bdq^{\prime}}
       +\frac{1}{2}\epsilon_{\bdq}
       \frac{\tilde{B}_{\bdq^{\prime}}}{\epsilon_{\bdq^{\prime}}}\Bigr).\label{eq:g-start}
\end{align}
Then,
in the long-wavelength limit for $D/J\ll 1$,
Eq. (\ref{eq:g-start}) reduces to
\begin{align}
  g_{\bdq}
  &\sim -\frac{1}{N}\sum_{\bdq^{\prime}}(q^{2}+q^{\prime 2})
  \frac{16J^{2}+4D^{2}}{4Jqq^{\prime}}
  +\frac{1}{N}\sum_{\bdq^{\prime}}\frac{q^{\prime}}{q}(4J+\frac{D^{2}}{2J})
  +\frac{1}{2N}\sum_{\bdq^{\prime}}\frac{q}{q^{\prime}}(4J+\frac{D^{2}}{2J})\notag\\
  &\sim -\frac{2}{N}J\sum_{\bdq^{\prime}}\frac{q}{q^{\prime}}\notag\\
  &=-2Jq\frac{q_{\textrm{c}}}{2\pi},\label{eq:g-last} 
\end{align}
where we have used
Eqs. (\ref{eq:e-limit}){--}(\ref{eq:B-limit}) and 
the replacement
$\frac{1}{N}\sum_{\bdq^{\prime}}\rightarrow \int_{0}^{q_{\textrm{c}}}\frac{dq^{\prime}}{2\pi}q^{\prime}$.
\end{widetext}

\end{document}